\documentclass[12pt,preprintnumbers,amsmath,amssymb]{revtex4}
\usepackage{graphicx}
\usepackage{color} 

\begin{document}

\title{Fractality in tumor growth at the avascular stage from a generalization of the logistic-Gompertz dynamics}

\author{Luiza M.~S. Miranda}
\thanks{Corresponding author: E-mail address: luizamayara16@hotmail.com}

\author{Andre M.~C. Souza}
\thanks{E-mail address: amcsouza@ufs.br}

\address{Departamento de Fisica, Universidade Federal de Sergipe, 49100-000, Sao Cristovao-SE, Brazil}

\date{\today}

\begin{abstract}
The dynamics between healthy and malignant cells at the avascular stage of growth is described by a set of chemical reactions representing the populations of both types of cells.  We obtain a generalization of the logistic-Gompertz dynamics and study the consequences in terms of the properties of tumor growth associated with the Warburg effect, mitosis and nutrient and space apoptosis. As a result of the stochastic nature of this process, a fractal morphology is observed in this region. Our results agree with experimental results for growth of tumor cell colonies.
\end{abstract}

\keywords{logistic map; avascular tumor growth; fractal edge}

\maketitle

\section{Introduction}
The dominant physical processes involved in the growth of cancerous tumors still pose a major challenge for science. Researches have observed the presence of characteristics associated with deterministic chaos, internal and external noise, dynamic behavior associated with nonlinear systems, relative robustness to internal and external perturbations and plasticity \cite{ara,sinh}. A mechanism for the dynamics of growth that encompasses all of these of complexity has not been described by any theory.
Several models of tumor growth have been introduced based on the assumption of only some of the above characteristics \cite{sinh,ende,roo,ampa,low,izq3,cel0}. Of course, it should be clear that the choice of some of the above characteristics must still be able to represent the growth of the tumor and certain aspects of its complexity, in order to make a solution possible.

Recently, a heuristic model based on simple chemical reactions was proposed by Izquierdo-Kulich et al.  to describe the dynamics of cancerous tumor growth at the avascular stage \cite{izq1}. The processes of mitosis and apoptosis and the action of the host are considered and the behavior is related to the logistic-Gompertz dynamics \cite{ara}.

In this work, we generalize the chemical reaction network proposed by Izquierdo-Kulich et al. \cite{izq1}. We also add the Warburg effect to our model, to take into consideration the fact that acidification of the medium of a normal cell can create a tumor \cite{war1,war2}. In the step that describes apoptosis, we separate the apoptosis based on nutrients and space processes and introduce a parameter $q$ that is associated with the density of the tumor cells.
The above processes do not exhaust all the mechanisms of tumor growth, since the complexity of the tumor surface, which is associated with the geometric fractal dimension of the contour, clearly shows that we cannot ignore the stochastic environment in which the tumor is found. We therefore also introduce a noise term.
A generalized logistic-Gompertz dynamics is presented
that includes these new components of the tumor growth. 

From the results obtained here, we develop a relation between the fractal structure of the cell and the parameter $q$ of our formalism.

The paper is organized as follows. In Sec. II, we introduce our chemical model of avascular tumor growth, and present some results derived from our model. In Sec. III, we develop a relation between the fractal structure of the cell and the parameter $q$ proposed in our formalism. Finally, Sec. IV presents our concluding remarks.

\section{The chemical model of avascular tumor growth}

Izquierdo-Kulich et al. proposed a heuristic model (henceforward referred to as the IRTN model) based on a network of simple chemical reactions \cite{izq1}. In the avascular phase, the tumor grows until it reaches a dormant state. If $A$ represents the normal cell population, $N$ is the number of proliferating tumor cells, $B$ is the host action, and $P$ represents a non-cancerous process, the tumor dynamics can be written in the form of the following chemical reactions:

(I) $N + A \rightarrow 2N$

(II) $2N \rightarrow P$

(III) $N + B \rightarrow P$

Step I is related to the process of mitosis and the reaction is associated with the mitosis constant $K_{m}$. Steps II and III correspond to apoptosis with the constant $K_{a}$ and host cell action with the constant $K_{b}$, respectively.
It is easy to see that the dynamic behavior of the system is given by the following logistic-Gompertz equation:
\begin{equation}
\label{gel}
\frac{dN(t)}{dt}= \gamma' N(t) - k_{a}N(t)^{2},
\end{equation}
where $\gamma' = Ak_{m}-Bk_{b}$. 
It is important to carry out a dimensional analysis of the model parameters. $N$, $A$ and $B$ are dimensionless, $t$ has the dimension of $[time]$, and the constants $k_{a}$, $k_{b}$, $k_{m}$ and $\gamma'$ have the dimension of $[time]^{-1}$. By implicitly setting $k_{b}=1$, we have assumed a timescale of $(k_{b})^{-1}$. 
The solution to the temporal dynamics of the number of tumor cells is
 \begin{equation}
\label{solg}
N(t)=\frac{\gamma' N(0)}{k_{a}N(0)(1-e^{-\gamma' t})+\gamma' e^{-\gamma' t} }.
\end{equation}
For $\gamma' > 0$ the stationary state has $N(t \rightarrow \infty) =\gamma' /k_{a}$ whereas for $\gamma' < 0$ we have $N(t \rightarrow \infty)=0$. 
A bifurcation occurs at $\gamma'=0$, which act as a second-order phase transition \cite{izq1,izq2}, consistent with the result of Prigogine and Lefever \cite{pri}.

Tumor cells not only differ from normal cells in terms of their appearance, but their entire energy metabolism is reassembled. The energy consumption of metabolic activities in normal cells depends mainly on mitochondrial oxidative phosphorylation, which is efficient and generates more adenosine triphosphate (ATP) than glycolysis. 
However, one of the metabolic characteristics of tumor cells is that they mainly convert glucose through aerobic glycolysis. This pathway for energy production in tumor cells is known as the Warburg effect \cite{war1,war2}. In the present work, we improve the IRTN model of this effect through the addition of a new step for the chemical equations.
Finally, in the step that describes apoptosis, we separate the apoptosis based on the nutrient and space processes, to allow the generalized model to be written in the form:

(I) $A + N \rightarrow  2N$

(II) $A \rightarrow N$

(III) $qN \rightarrow P$

(IV) $N + B \rightarrow P$.

The mitosis process is described in Step I, in which there are $N$ tumor cells and $A$ normal cells. Step II describes the Warburg effect, in which a normal cell can become a tumor following acidification of the medium.
Step III describes the process of apoptosis by nutrients, where for this to happen, there must be a density of tumor cells $q$. Step IV describes apoptosis by space, in which $B$ defense cells of the host tissue generate apoptosis.
The new dynamic behavior of the system is given by the following logistic-Gompertz equation:
\begin{equation}
\label{dina}
\frac{dN(t)}{dt}= \gamma N(t) - k_{an}N(t)^{q} + Ak_{mw},
\end{equation}
where $\gamma = Ak_{m}-Bk_{ae}$. We have assumed that the timescale is $(k_{ae})^{-1}$. 
We take into consideration two processes of mitosis. The first term refers to the duplication of cancer cells at a constant rate $k_{m}$. The possible transformation of normal cells into cancer cells linked to the Warburg effect takes place at a constant rate $k_{mw}$. In this case, the acidity of the medium, favored by the anaerobic respiration of cancer cells, causes fermentation, and therefore serves as a catalyst for the appearance of new cancer cells.

Apoptosis is linked to the mechanisms for the destruction of cancer cells by the body's defense systems. We can divide these mechanisms into two types: apoptosis by nutrients at a constant rate $k_{an}$, and apoptosis by space, with a constant rate $k_{ae}$.
Nutrient apoptosis involves the mechanisms of competition for nutrients between cells, in which the high density of cancer cells in each region of the tissue favors a lack of nutrients and thus their destruction.
However, the lack of nutrients does not fully describe some basic features of tumor growth. Cell proliferation is greatly inhibited inside the colony
or tumor \cite{bru1}, and the geometric and dynamic properties of cell colony growth are affected by external pressure \cite{brup}.
In view of this behavior, an expression for tumor growth cannot be extracted considering only apoptosis by nutrients, and it is therefore of fundamental importance to add apoptosis by space as a mechanism for inhibiting tumor growth. In the present model, this mechanism is introduced with a rate reaction $k_{ae}$, and is proportional to the density of cancer cells in the region, represented by the parameter $q$.
The IRTN model is obtained for $k_{mw}=0$, $k_{ae}=1$, $k_{an}=k_{n}$, $\gamma=\gamma'$ and $q=2$.

We emphasize that a similar parameter $q$ was introduced in a previous work \cite{tsacel}, as part of a model based on the ideas of nonextensive statistical mechanics \cite{tsa1,tsa2}.
From different values of the nonextensive parameter $q$, different types of
tumor evolution were obtained\cite{tsacel}.
Based on the ideas underlyng nonextensive entropy, the study in \cite{tsacel} showed that tumor growth depends on the nonextensive parameter $q$: the smaller the value of $q$, the more aggressive the tumor becomes. It is important to note that although the parameter $q$ used in the present work has similarities with that used in \cite{tsacel}, it is not the same, since the equations for dynamics are not the same in both works, and the Warburg effect and stochastic noise are not considered in \cite{tsacel}.

These processes are not likely to be an exhaustive description of all the mechanisms of tumor growth, since the complexity of the tumor surface, which isassociated with the geometric fractal dimension of the contour, clearly shows that we cannot ignore the stochastic environment in which the tumor is found \cite{bruprl}. In addition to the processes of mitosis and apoptosis, we consider a noise term $\eta_i$ taken
out of a box uniform distribution within $[-W/2,W/2]$, where $W$ is the disorder strength and $i$ the sample.

Given the above, the dynamics of tumor growth can be described by the following expression
\begin{equation}
\label{geral}
\frac{dN(t)}{dt}= \gamma N(t) - k_{an}N(t)^{q} + Ak_{mw} + \eta_i.
\end{equation}

\begin{figure}
	\centering
	\includegraphics[width=8cm,angle=0]{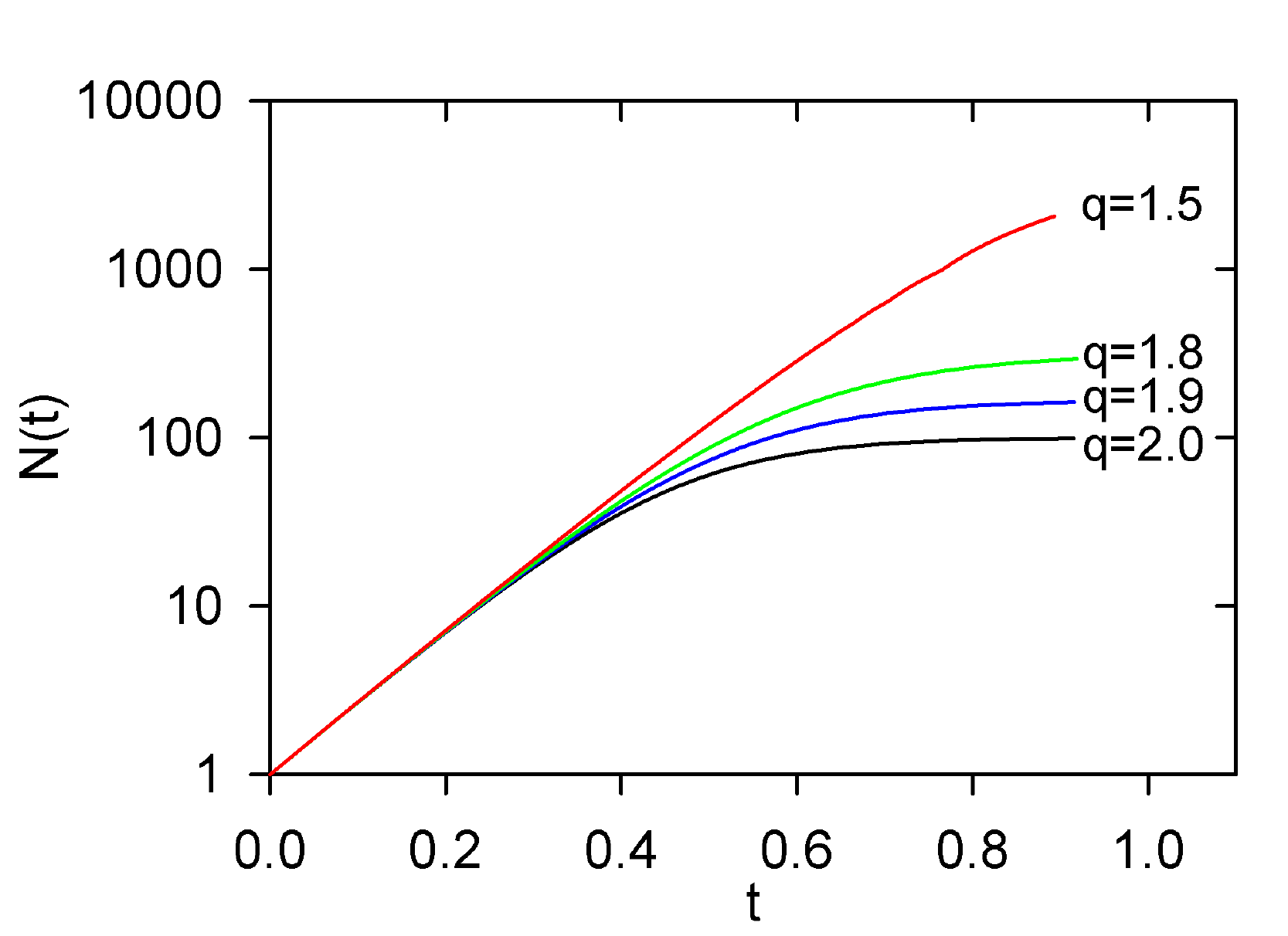}
	\caption{Time evolution of tumor cells $N(t)$ in the absence of the Warburg effect ($k_{mw}=0$) and a noise term($\eta=0$). We use the timescale as $(k_{ae})^{-1}$.
	We consider $q=1.5$ (red curve), $1.8$ (green curve), $1.9$ (blue curve) and $2.0$ (black curve).}
	\label{fig1}
\end{figure}

Fig. \ref{fig1} shows the evolution of the number of cancer cells as a function of time for $q=1.5$,$1.8$,$1.9$ and $2.0$.
The Warburg effect and noise are not considered, i.e., $k_{mw}=0$ and $\eta=0$. In this limit, Eq. \ref{geral} becomes a Bernoulli differential equation, and it is therefore to show that
 \begin{equation}
\label{solsw}
N(t)=\frac{\gamma^{\frac{1}{q-1}} N(0)}{\left[ k_{an}N(0)^{q-1}(1-e^{-(1-q)\gamma t})+\gamma e^{-(1-q)\gamma t} \right]^{\frac{1}{q-1}}}.
\end{equation}
The tumor grows to a certain size and then goes into a dormant state for every value of $q$.
An arbitrary value of $q$ does not change the universality class of the IRTN model ($q=2$). A bifurcation occurs at $\gamma=0$ and for $\gamma > 0$ the stationary state has $N(t \rightarrow \infty) =(\gamma /k_{an})^{1/(q-1)}$, while for $\gamma < 0$ we have $N(t \rightarrow \infty)=0$.
Considering that $q$ is directly linked to the density of tumor cells so that there is apoptosis, we can see that the higher the density of the tumor cells, the faster the tumor stabilizes. This is because a lack of nutrients is associated with the mechanisms of competition for nutrients between cells. Thus, the high density of tumor cells in each region of the tissue favors a lack of nutrients and thus their destruction. Tumor growth is restricted by the ability of the tissue to transport nutrients.
Note that in the initial stage of tumor growth the role of $q$ is almost imperceptible.
As noted previously in \cite{tsacel}, the smaller values of $q$, the more aggressive the tumor becomes.
Since the number of infected cells $N(t)$ in the steady state becomes larger as $q$ decreases.

\begin{figure}
	\centering
	\includegraphics[width=8cm,angle=0]{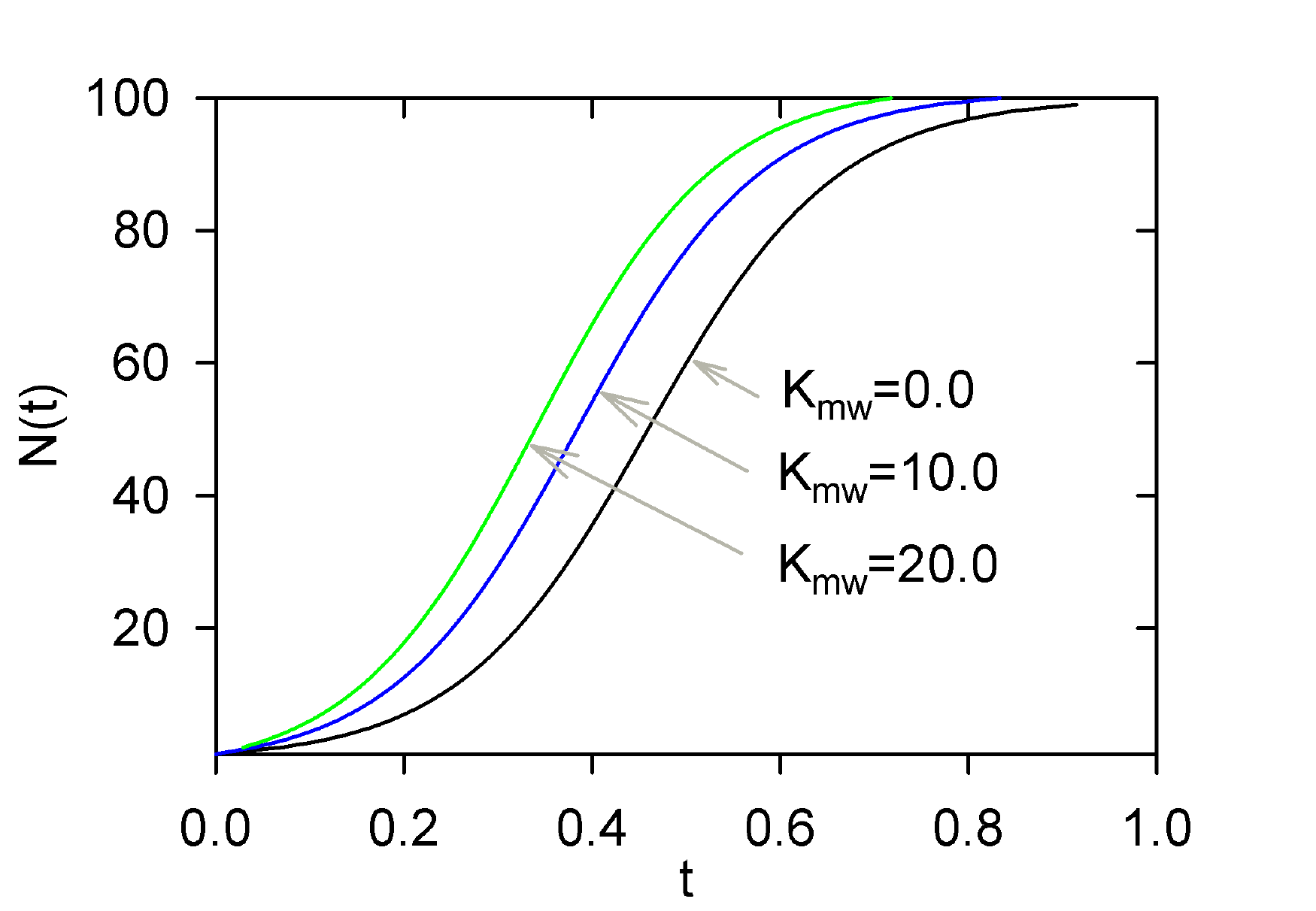}
	\caption{Time evolution of tumor cells $N(t)$ in the absence of a noise term($\eta=0$) and $q=2$, for Warburg rates $k_{mw}=0$ (black curve), $10.0$ (blue curve) and $20.0$ (green curve). We use a timescale $(k_{ae})^{-1}$.}
	\label{fig2}
\end{figure}

The Warburg effect is shown in Fig. \ref{fig2} for a fixed value of $q=2$. The analytical solution for the dynamics of tumor growth is
 \begin{equation}
\label{solcw}
N(t)=\frac{ N(0)\left[\Delta + \gamma + (\Delta-\gamma)e^{-\Delta t}\right] + 2k_{mw} (1-e^{-\Delta t})}{2k_{an}N(0)(1-e^{-\Delta t})+(\Delta+\gamma)e^{-\Delta t} +\Delta-\gamma },
\end{equation}
where $\Delta=\sqrt{\gamma^{2}+4k_{an}k_{mw}}$.
it can be seen from Fig. \ref{fig2} that with an increase in the Warburg effect constant $k_{mw}$, tumor growth tends to become faster. This is because cellular respiration by aerobic glycolysis is faster despite being an inefficient means of ATP production. In the Warburg effect, tumor cells conduct metabolism mainly by aerobic glycolysis, producing lactate even in the presence of oxygen and fully functioning mitochondria. This process increases the intracellular pH and acidifies the extracellular environment, turning normal cells into tumor cells. This metabolic characteristic makes the tumor more robust and aggressive. Similar results are obtained for the other cases of $q$.

\begin{figure}
	\centering
	\includegraphics[width=8cm,angle=0]{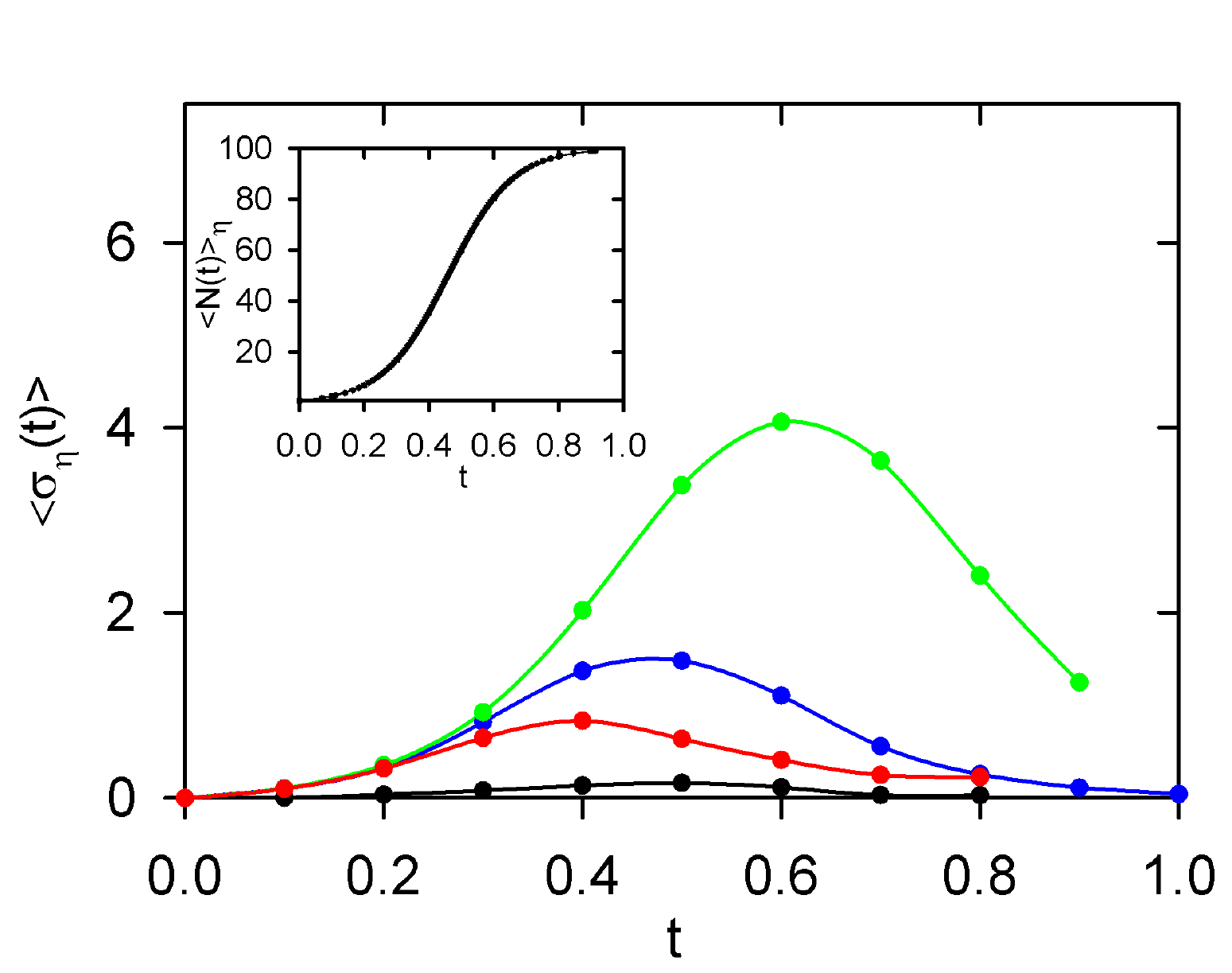}
	\caption{Time evolution of the standard deviation of tumor cells $\langle \sigma(t) \rangle_{\eta}$. The curves are: black color ($q=2$, $k_{mw}=0$, $W=0.1$); red color ($q=2$, $k_{mw}=10.0$, $W=0.1$); blue color ($q=2$, $k_{mw}=0$, $W=1.0$); green color ($q=1.8$, $k_{mw}=0$, $W=1.0$). 
	We use a timescale $(k_{ae})^{-1}$ and $m=100$ samples. INSET:  $\langle N(t) \rangle_{\eta}$ versus $t$ for $q=2$, $k_{mw}=0.0$ and $W=0$ (circles), $W=0.1$ (squares), $W=1.0$ (triangles). Differences in the curves are not noticeable at this scale.}
	\label{fig3}
\end{figure}

For the case where $\eta \ne 0$, the quantities can be determined by their average $\langle ... \rangle_{\eta}$ over $m$ independent realizations of the noise term $\eta$.
We find that the noise does not change the size of the tumor.
From this point, we have explored the model solutions numerically.
We obtain the result $\langle N(t) \rangle_{\eta} \approx \langle N(t) \rangle_{\eta=0} = N(t)$. The inset to Fig \ref{fig3} shows this result, in which the time evolution of $\langle N(t) \rangle_{\eta}$ is similar for three different values of $W$.

The noise effect is found by analyzing the standard deviation of the variables. For $\langle N(t) \rangle_{\eta}$, the standard deviation is
\begin{equation}
\label{sigm}
\langle \sigma(t) \rangle_{\eta}= \sqrt{\langle N(t)^{2} \rangle_{\eta}-\langle N(t) \rangle_{\eta}^{2}}.
\end{equation}

Fig. \ref{fig3} shows $\langle \sigma(t) \rangle_{\eta}$ for typical values of the model parameters. 
We consider $m=100$ samples to noise term $\eta_i$ average.
It is possible to see that the standard deviation with stochastic noise increases during tumor growth and then decreases when the tumor enters the dormant state.
Of course, increasing the disorder strength $W$ also increases the deviation $\langle \sigma(t) \rangle_{\eta}$.
For fixed values of the other parameters of the model, we find that the standard deviation increases as $q$ decreases, whereas it increases as $k_{mw}$ increases.

\begin{figure}
	\centering
	\includegraphics[width=8cm,angle=0]{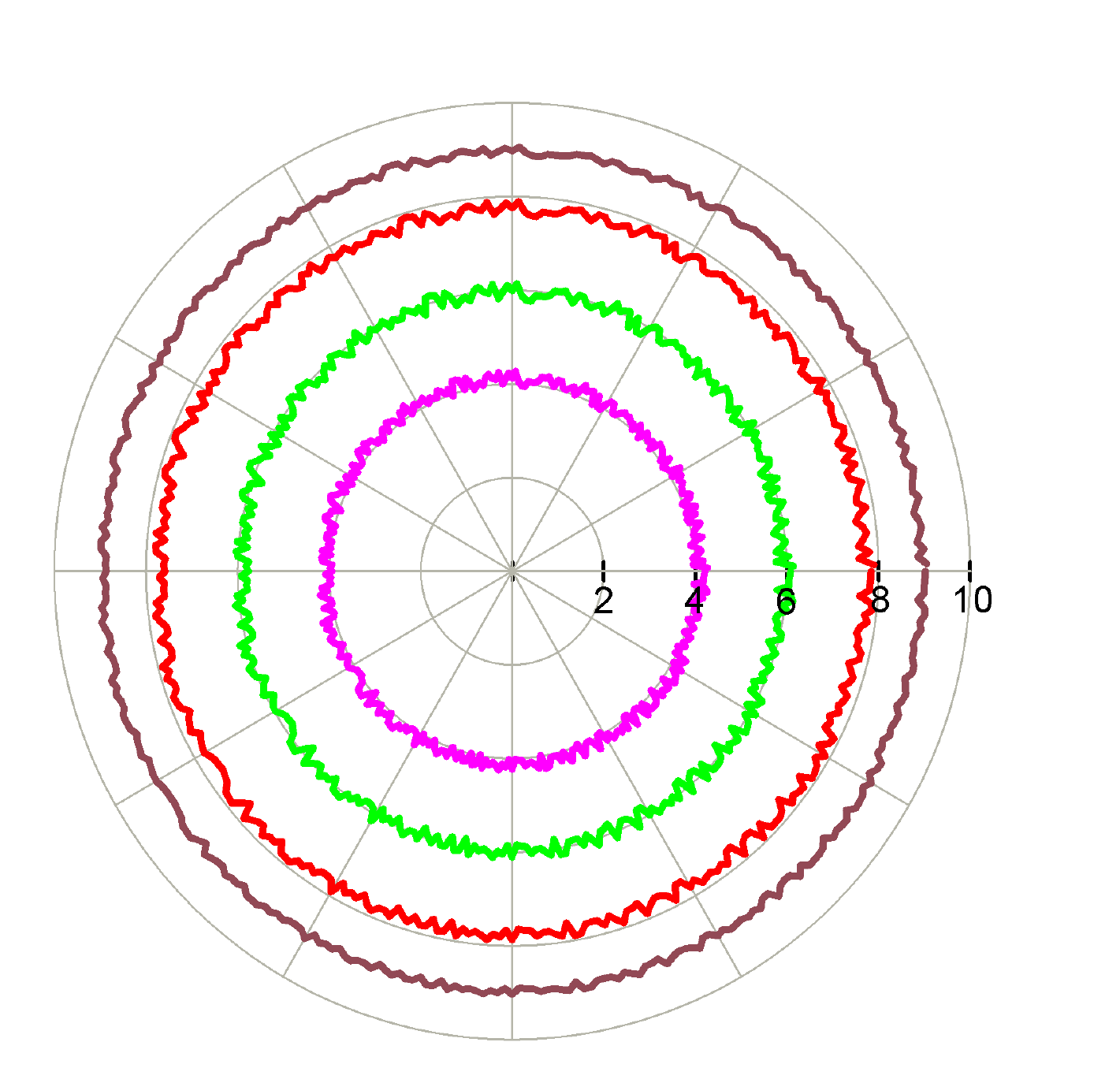}
	\caption{Tumor interface versus radius $r(t)/R$ at times $t=0.2$ (pink curve), $0.4$ (green curve), $0.6$ (red curve), and $0.8$ (brown curve) for $q=2$, $k_{am}=0.1$, $k_{mw}=0$, $\gamma=10.0$ and $W=1.0$.}
	\label{fig4}
\end{figure}
\begin{figure}
	\centering
	\includegraphics[width=8cm,angle=0]{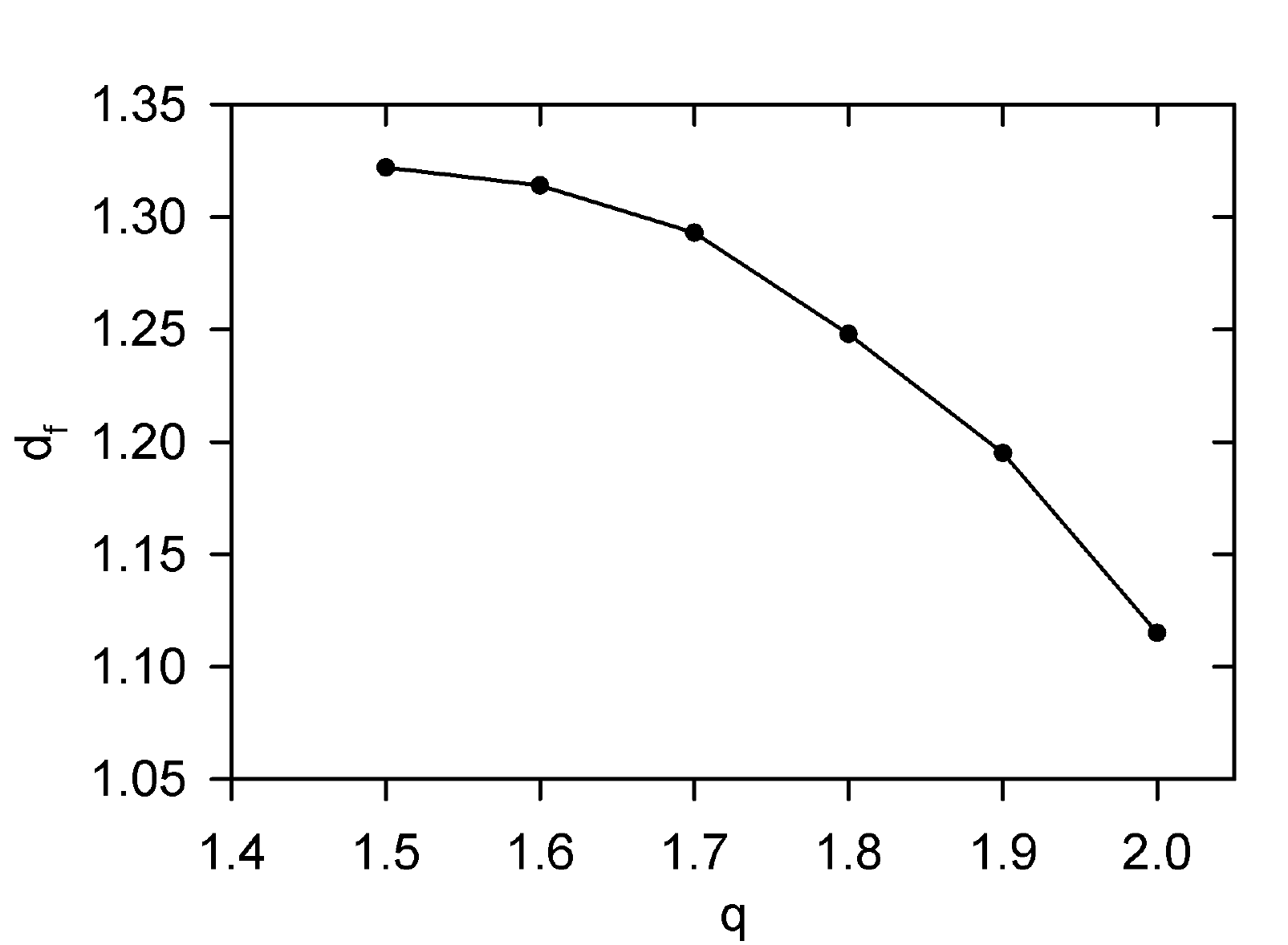}
	\caption{Fractal dimension of the tumor interface calculated with the
		box-counting method at time $t$, corresponding to $\langle r(t) \rangle_{\eta} \approx 9R$ as a function of $q$. We use the parameters $k_{am}=0.1$, $k_{mw}=0$, $\gamma=10.0$ and $W=1.0$.}
	\label{fig5}
\end{figure}

\section{Fractal structure of a tumor cell}
Experiments performed in mice have shown that the dynamics of tumor growth create a host-tumor interface with a rough structure \cite{bruprl}. There is even a change in
the growth dynamics of cell colonies that is induced by a change in pressure \cite{brup}.
Considering the surface roughness structure of the tumor, it is possible to determine the scale invariance of the fractal interface from the calculation of its fractal dimension $d_{f}$ \cite{bru1}.

Different cell lines and different types of tumors were studied and fractal dimensions were obtained for a range of $d_{f} \in [1.05, 1.35]$. A direct relationship between the fractal dimension and the type of tumor was not conclusive. 
However, it became very clear that there was a need for further studies to relate these features of the tumor and to some clinical strategies \cite{bru1}.
Within this context, we present an analysis that relate the fractal dimension of the tumor surface and to the components of the formalism introduced in this work.
Our main hypothesis is the assumption that the surface radius of the tumor $r(t)$ is proportional to the square root of the number of tumor cells $N(t)$.
We consider that the tissue that forms the cellular substrate is two-dimensional, and hence $r(t) =R \sqrt{N(t)}$. The different radius values are obtained from different realizations of the noise term $\eta_i$.

In Fig. \ref{fig4}, we see the tumor contours as a function of the radius $r(t)/R$ for different times. The times are $k_{ae}t=0.2$, $0.4$, $0.6$ and $0.8$, which were chosen based on the growth dynamics shown in previous figures. We used values of $q=2$, $k_{am}=0.1$, $k_{mw}=0$, $\gamma=10.0$ and $W=1.0$, and $m=360$ samples (where each represented one of the $360$ angles in the figure).
Similar results were found for arbitrary values of $q$.
We performed the same analysis for different numbers of angles, and the results obtained for the fractal dimension were the same, indicating that the number of samples does not affect the roughness structure of the dynamics in the present model.

The box-counting method was used to calculate the fractal dimension $d_{f}$.
Fig. \ref{fig5} shows the fractal dimension of the tumor at time $t$, corresponding to $\langle r(t)/R \rangle_{\eta} \approx 9$ as a function of $q$. We use the parameters $k_{am}=0.1$, $k_{mw}=0$, $\gamma=10.0$ and $W=1.0$. 
It is straightforward to observe that the value of $d_{f}$ are within the interval obtained from the experimental results under normal pressure conditions \cite{bru1}.
We find that increasing $q$ decreases $d_{f}$, since the value of $q$ indicates the density of the tumor cells in the process of apoptosis in dispute for nutrients. We find that an increase in $d_{f}$ indicates a lower density of cells in dispute for nutrients, this is because the higher the density of malignant cells, the faster the tumor stabilizes, as the standard deviation $\langle \sigma(t) \rangle_{\eta}$ is smaller. 
We found that for tumor growth, $d_{f}$ initially increases until it reaches a maximum, within an intermediate evolution time, and then decreases as the tumor enters the steady state.

\section{Conclusions}
We propose a model for tumor growth at the avascular stage based on a new chemical reaction network, in which we consider the process of mitosis by cell division, the Warburg effect, and apoptosis by nutrients and space. In addition, we include the stochastic noise, and fluctuations at the avascular stage of the tumor are observed.
From the logistic-Gompertz dynamics, we see that at the outset, tumor growth is fast, whereas over time the cell population saturates and reaches a dormant state. The time to reach a steady/dormant state depends on the index $q$, which represents the density of tumor cells. The Warburg effect clearly affects tumor growth, as tumor cells conduct their metabolism mainly by aerobic glycolysis. This metabolic characteristic makes cancer more robust and aggressive, due to acidification of the extracellular environment. The relationship between the fractal dimension of the tumor surface and the tumor cell density $q$ is implicit in the dynamics of the model.
It is important to mention that the results obtained here agree with the experimental results recently presented for the growth of tumor cell colonies \cite{bru1,brup}.
Our results suggest that the fractal dimension of the tumor's borders increases to a maximum value in a certain intermediate time of its evolution, and then decreases as the tumor moves towards its stationary state. Experiments to test this hypothesis are welcome.

\section*{ACKNOWLEDGMENTS}
We acknowledge the support of the CNPq (Brazilian agency).

\end{document}